# Low-Threshold Degenerate Optical Parametric Oscillations in Bichromatically-Pumped Normal-Dispersion Photonic-Crystal Microresonator

Valery E. Lobanov[1], Nadezhda S. Tatarinova[1,2], Artem E. Shitikov[1], Olga V. Borovkova[1], Igor A. Bilenko[1,3], and Dmitry A. Chermoshentsev[1,2]

[1]Russian Quantum Center, 143026 Skolkovo, Russia
[2]Moscow Institute of Physics and Technology, 141701 Dolgoprudny, Russia
[3]Faculty of Physics, Lomonosov Moscow State University, 119991 Moscow, Russia

**The process of excitation of degenerate optical parametric oscillations via bichromatic pump is studied numerically in normal-dispersion photonic-crystal microresonator. It is demonstrated that the photonic-crystal structure with two split modes placed symmetrically at the particular interval from the pumped modes provides significant reduction in pump power threshold for the considered process. The parameter range for this phenomenon is determined. Introduction of mode splitting at the signal mode located in the center between pumped modes leads to an increase in the generation threshold.**

## 1. Introduction

The use of modern microresonator structures with high Q factors and ultra-compact size creates favorable conditions for the implementation of various nonlinear optical processes, including frequency conversion, generation of frequency combs and dissipative solitons, making such systems an ideal platform for nonlinear photonics [1-15]. The capabilities of microresonator photonics have further expanded in recent years with the development of methods of nanostructured or photonic-crystal microresonators fabrication [16-33]. It has been demonstrated that periodic modulation of the microring parameters allows to create a bandgap for a specific mode or number of modes, leading to the mode splitting and effective mode shift [16-18]. In turn, such a controlled mode shift allows to satisfy phase-matching conditions required for efficient realization of particular nonlinear processes. For example, in photonic-crystal microresonators with anomalous group velocity dispersion (GVD) it has been demonstrated the efficient soliton microcomb generation for various pump schemes [19,20,27], excitation of dark-to-bright pulse continuum [21] and the enhancement of optical parametric oscillations [23-25]. Besides that, the use of such microresonators has significantly expanded the possibilities of realization of different nonlinear processes in spectral ranges with normal GVD that may be a challenging task in conventional microresonator without photonic-crystal structure [34, 35]. Namely, efficient generation of dark solitons or so-called platicons [20, 28, 29, 32, 33] and optical parametric oscillations [30,31] in the normal dispersion regime have been reported.

Among other nonlinear processes, the process of degenerate optical parametric oscillations (DOPO) is of particular interest due to its applications in coherent computing, quantum computing, and random number generation [36-39]. Efficient realization of DOPOs have been demonstrated in media with either second-order $\chi^{(2)}$ or third-order $\chi^{(3)}$ nonlinearities [40–44]. One of the techniques for DOPO realization in microresonators with Kerr nonlinearity is based on the degenerate four-wave mixing process upon bichromatic (or dual-frequency) pump [45, 46]. The generated signal exhibits phase bistability that is a crucial feature for implementing advanced computational schemes, such as all-optical coherent Ising machines [44, 47-49]. Furthermore, below the DOPO pump power threshold, the microresonator can operate as a generator of squeezed states of light [50]. Recent studies have shown the perspectives of using microresonators with normal GVD and symmetric frequency scan for efficient DOPO realization [46, 51, 52].

The threshold power reduction is an important task for practical applications of this process. Solving this problem will both diminish the impact of thermal and undesirable nonlinear effects and improve the energy efficiency of devices based on DOPO process.

In recent work [51], it has been demonstrated that controlled dispersion engineering – specifically, a shift of the signal mode and a symmetric shift of the particular side modes – can substantially reduce the DOPO generation threshold. However, the analysis has been performed using simplified model, not related to the particular system. Since necessary mode shifts can be realized in photonic-crystal microresonators, in our work we numerically study the process of degenerate optical parametric oscillations via bichromatic pump in normal-dispersion photonic-crystal microresonator. It is demonstrated that photonic-crystal structure with two split modes placed symmetrically at the particular interval from the pumped modes provides significant decrease of pump power threshold for the considered process. The parameter range for this phenomenon is determined. On the contrary, introduction of mode splitting at the signal mode located in the center between pumped modes leads to an increase in the generation threshold.

## 2. Model

We studied the process of DOPO in normal-dispersion microresonator via bichromatic pump and linear-in-time pump frequency scan taking backscattering into account. In our work, we used a model derived from the first principles in [53, 54] which considers the linear forward-backward wave coupling with the frequency-dependent coupling strength and nonlinear cross-action:

$$\begin{cases} \frac{\partial \psi_1}{\partial \tau} = i\frac{d_2}{2}\frac{\partial^2 \psi_1}{\partial \varphi^2} - [1 + i\alpha]\psi_1 + i\sum_{\mu} \beta_{\mu} a_{2\mu} \exp(i\mu\varphi) + \\ +i\left( |\psi_1|^2 + 2\frac{U_2}{2\pi} \right)\psi_1 + 2F\cos(n\varphi), \\ \frac{\partial \psi_2}{\partial \tau} = i\frac{d_2}{2}\frac{\partial^2 \psi_2}{\partial \varphi^2} - [1 + i\alpha]\psi_2 + i\sum_{\mu} \beta_{\mu} a_{1\mu} \exp(-i\mu\varphi) + i\left( |\psi_2|^2 + 2\frac{U_1}{2\pi} \right)\psi_2. \end{cases} \quad (1)$$

Here $\psi_1$ and $\psi_2$ are the slowly varying envelope of the forward and backward wave in microresonator, respectively ($\psi_1(\varphi) = \sum_{\mu} a_{1\mu} \exp(i\mu\varphi)$, $\psi_2(\varphi) = \sum_{\mu} a_{2\mu} \exp(-i\mu\varphi)$, where $a_{1\mu,2\mu}$ are $\mu$-th forward and backward mode amplitudes, $\mu$ is the mode number, calculated from the signal mode), $\tau = \kappa t / 2$ denotes the normalized time, $\kappa = \frac{\omega_0}{Q}$ is the cavity decay rate, $Q$ is the loaded quality factor, $\omega_0$ is the signal mode resonant frequency, $\varphi \in [-\pi; \pi[$ is an azimuthal angle in a coordinate system rotating with the angular frequency equal to the microresonator free spectral range (FSR) $D_1$, $d_2 = 2D_2 / \kappa$ is the GVD coefficient, positive for the anomalous GVD and negative for the normal GVD. The microresonator eigenfrequencies are assumed to be $\omega_{\mu} = \omega_0 + D_1\mu + \frac{1}{2}D_2\mu^2$. We assumed that the modes with numbers $\mu = +n$ and $\mu = -n$ are pumped, and the signal is generated between them at $\mu = 0$. $\alpha = \frac{2}{\kappa}\left(\omega_0 - \frac{\omega_{p_1} + \omega_{p_2}}{2}\right)$ is the normalized detuning of the signal frequency, equal to the half-sum of the pump frequencies $\omega_{p_1}$ and $\omega_{p_2}$, from the signal mode resonance $\omega_0$. The normalized pump amplitude is $F = \sqrt{\frac{8\omega_p c n_2 \eta P_{in}}{\kappa^2 n_0^2 V_{eff}}}$, where $n_0$ is the refractive index of the pumped microresonator modes, $c$ is the speed of light in vacuum, $V_{eff}$ is the effective mode volume, $n_2$ is the nonlinear refractive index, $\eta$ is the coupling efficiency [$\eta = 1/2$ for critical coupling, $\eta \to 1$ for overcoupled regime], $P_{in}$ is the input pump power assumed to be equal for each pump source, $U_{1,2} = \int |\psi_{1,2}|^2 d\varphi$ is forward/backward wave power; $\beta_\mu$ is the normalized $\mu$-th mode backscattering coefficient [equal to the ratio of the linear $\mu$-th mode splitting value and loaded microresonator linewidth $\kappa$]. We considered the case of the photonic-crystal microresonator when splitting coefficient is not the same for all modes that can be realized by particular modulation of microring parameters, i.e. its width.

We integrated Eqs. (1) numerically using split-step Fourier algorithm with periodic boundary conditions, considering non-uniform linear mode coupling in the spectral domain. Spatial grid of 512 points was used. It was checked that a further increase in the number of grid points did not change the results. To account for vacuum fluctuations in this system, we periodically added a weak noise-like seed with a mean amplitude of $10^{-6}$ during each simulation.

## 3. Degenerate optical parametric oscillations with backscattering

For numerical analysis of the conditions of the degenerate optical parametric oscillations process, we considered the case of normal dispersion ($d_2 < 0$) and symmetric pump frequency scan which was shown to be the efficient method of the realization of DOPOs [51, 52]. It means that modes at $\mu = +n$ and $\mu = -n$ were pumped identically, with the same normalized pump amplitude $F$ and the same offset of each pump frequency from the corresponding mode throughout the scan process.

First, we considered the case where the linear coupling between the forward and backward waves was the same for all modes: $\beta_\mu = \beta^{bg}$ [see the scheme of the process in Fig. 1]. Using Eqs. (1) we found numerically the threshold pump amplitude $F_{thr}$ of the degenerate optical parametric oscillations for different values of the dispersion parameter $-n^2 d_2$. The threshold pump value is determined by monitoring the intracavity power of the signal mode and identifying the conditions at which it exceeds the initial noise level [51]. It was revealed that for small values of the background backscattering coefficient $\beta^{bg}$ such dependence almost coincided with that obtained in [51] (see Fig. 2). Threshold pump value was found to have minimum at the particular value of the dispersion parameter [approximately, $F_{thr} \approx 1.24$ at $-n^2 d_2 \approx 2.4$ and $\beta^{bg} = 0.01$]. Interestingly, the optimal parameters are independent of the pump interval $2n$, with the dispersion parameter $-n^2 d_2$ remaining constant.

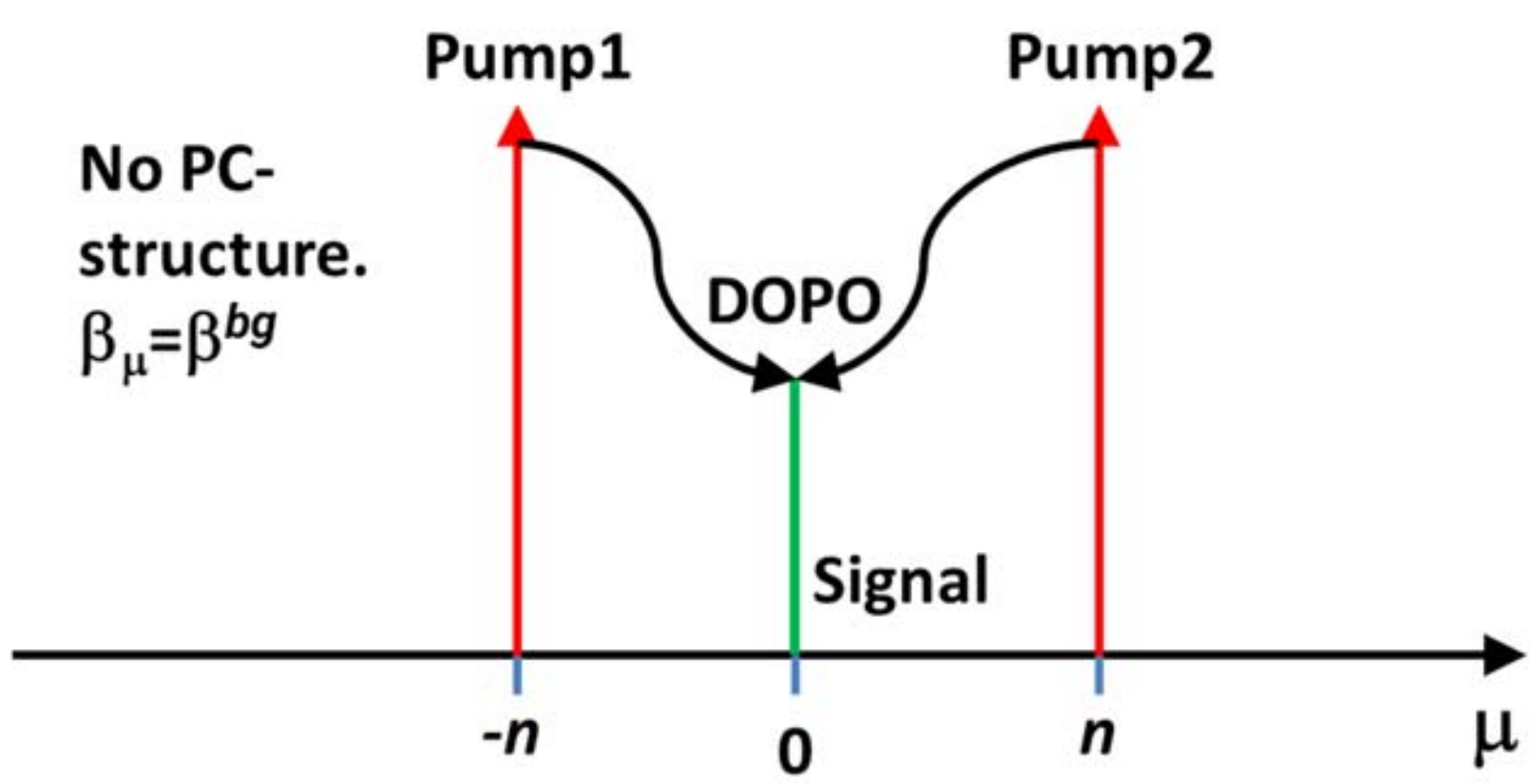

**Fig. 1.** Scheme of the process of DOPO at bichromatic pump in the presence of backscattering in conventional microresonator without photonic-crystal (PC) structure ($\beta_\mu = \beta^{bg}$).

We studied in more detail all possible generation regimes arising upon linear-in-time frequency scan $\alpha(\tau) = \alpha(0) + v\tau$ for $n = 5$. Such value of $n$ can provide dispersion parameter $-n^2 d_2$ close to the optimal value for experimentally feasible values of $d_2$. The scan rate $v$ was chosen small enough ($v = 0.0005$) to reveal all possible effects arising upon pump frequency scan. Simulation results are summarized in Fig. 2.

The generation of the desired signal at $\mu = 0$ occurs above some threshold pump value in area 1 in Fig. 2 [see corresponding spectrum in Fig. 3***(a)***]. It should be noted that the considered DOPO process is always accompanied by a process of non-degenerate four-wave mixing, leading to the appearance of a frequency comb with a period equal to the frequency interval between pump lines. This additional process does not prevent parametric signal generation via DOPO but leads to an increase in the generation threshold [51]. As the dispersion parameter increases, in addition to the excitation of the central mode, other regimes also appear even at lower pump amplitudes. In area 2 (Fig. 2) the result is non-deterministic, and for the same parameters but different noise-like initial conditions, either one central mode ( $\mu = 0$ ) [Fig. 3***(a)***], or two side modes ( $\mu = \pm 1$ ) [Fig. 3***(b)***], or more complex signals with at least three components ( $\mu = 0, \pm 1$ ) [Fig. 3***(c)***] can be generated between pumped modes. At larger values of the dispersion parameter, another region appears [area 3 in Fig. 2] where, at smaller values of the pump amplitude, deterministic generation occurs in several central modes [at least at $\mu = 0, \pm 1$, see Fig. 3***(c)***]. In area 4 in Fig. 2 deterministic excitation of side modes takes place ( $\mu = \pm 1$ ). Note, that obtained results are in a good agreement with the analysis of the parametric gain profile performed in [44] for the system of 3 or 4 interacting modes. We also checked that property of phase-bistability of the generated DOPO signal did not disappear in the presence of backscattering. Note, that phase bistability is also present at backward wave at signal frequency. The power of this wave increases with backscattering level.

The features of the analyzed processes do not change with increasing background backscattering level defined by $\beta^{bg}$ (compare panels ***(a)***, ***(b)*** and ***(c)*** in Fig. 2 obtained for different values of $\beta^{bg}$ ). However, one may notice that with the growth of $\beta^{bg}$ pump threshold for the deterministic generation of the desired central mode also increases and additional generation regimes appear at smaller values of the dispersion parameter [see Fig. 2***(d)***].

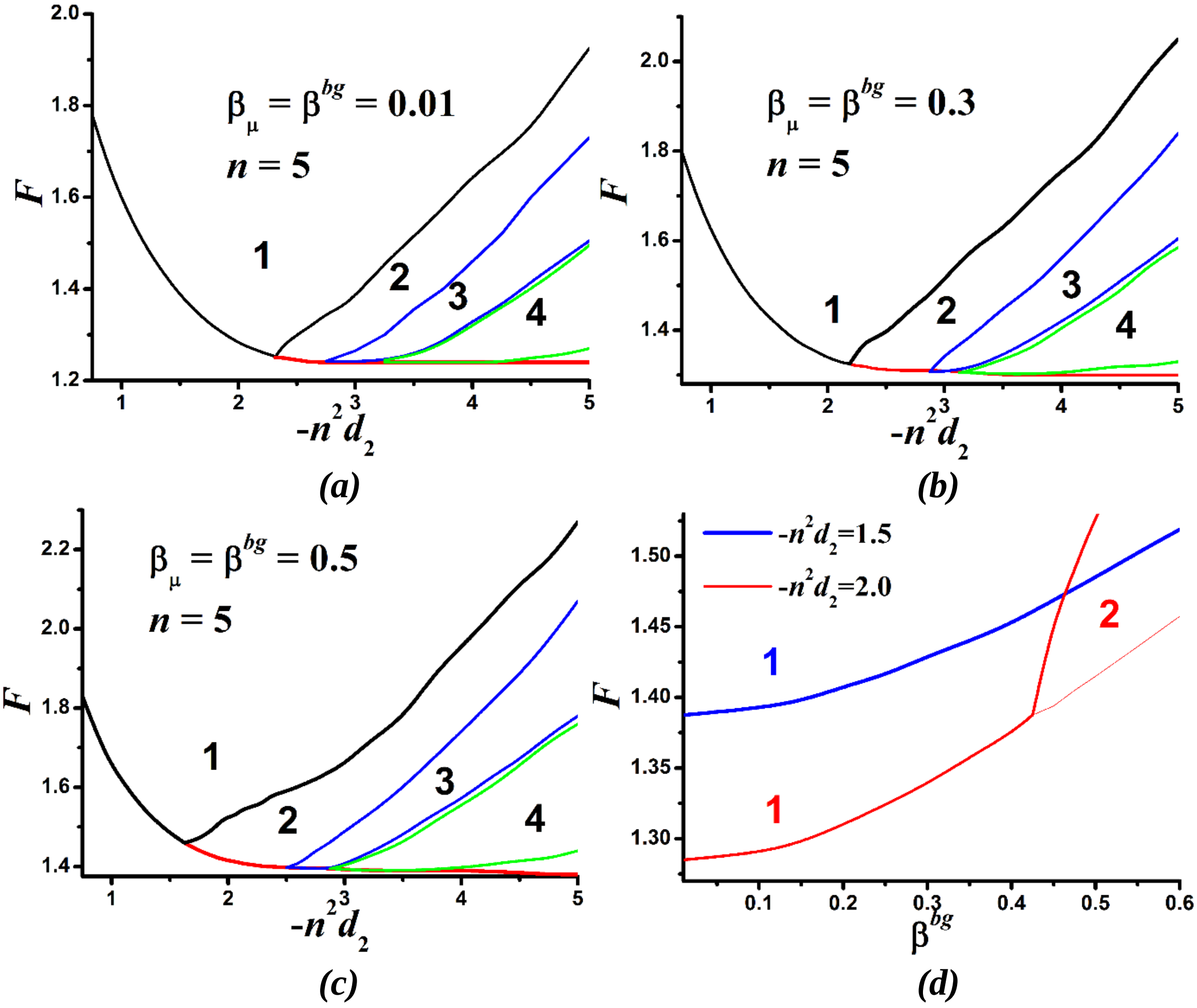

**Fig. 2.** Map of generation regimes arising in bichromatically-pumped normal dispersion microresonator upon linear-in-time symmetric pump frequency scan. Area 1 corresponds to the generation of the central mode ($\mu = 0$), area 2 – to the non-deterministic result depending on the particular realization of the noise-like input conditions, area 3 – to the generation of several modes (at least three central modes at $\mu = 0, \pm 1$) between pumped modes, area 4 – to the generation of side modes ($\mu = \pm 1$). In all cases $n = 5$, $v = 0.0005$. All quantities are plotted in dimensionless units.

It is worth noting that in all cases, starting from a certain value of the dispersion parameter $-n^2 d_2$, the threshold for generating frequency components between the pump lines ceased to depend on the dispersion value [see red line in Fig. 2***(a)*** – ***(c)***]; at the same time, the threshold for the process of deterministic generation of the central line at $\mu = 0$ increased rapidly with increasing dispersion parameter [see black line in Fig. 2***(a)*** – ***(c)***].

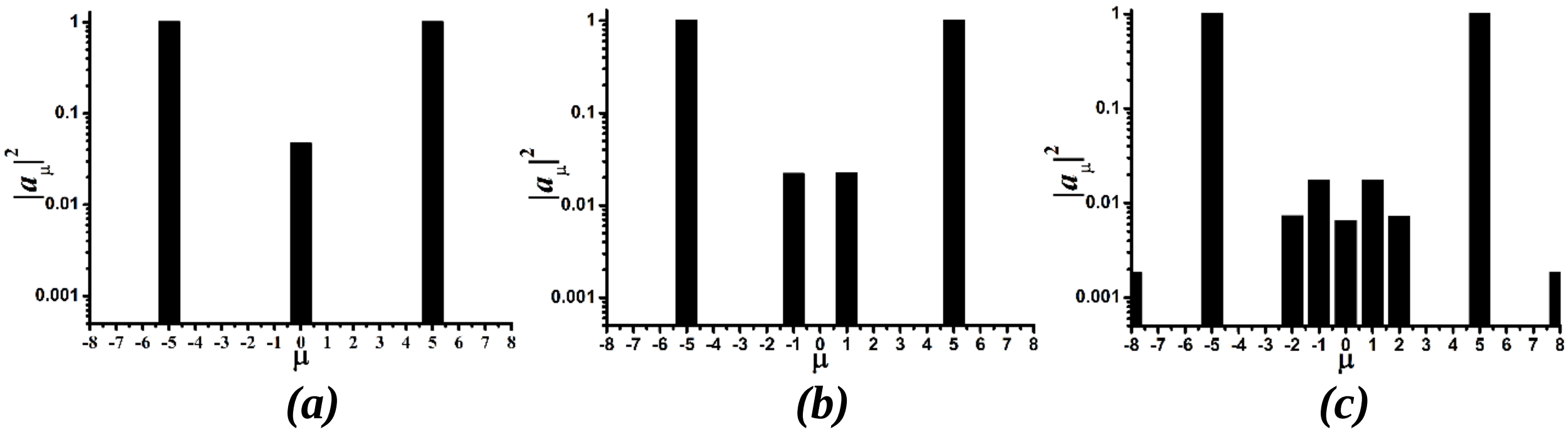

**Fig. 3.** Spectra obtained via linear-in-time symmetric frequency scan of bichromatical pump in normal dispersion microresonator with backscattering. Panel ***(a)*** corresponds to the generation of the central mode ($\mu = 0$), panel ***(b)*** – to the generation of side modes ($\mu = \pm 1$), panel ***(c)*** – to the generation of at least three central modes ($\mu = 0, \pm 1$) between pumped modes at $\mu = \pm 5$. In all cases 512 modes were considered at $n = 5$, $-n^2 d_2 = 2.5$, $F = 1.26$, $\alpha = 3.5$, $\beta^{bg} = 0.01$, $v = 0.0005$. All quantities are plotted in dimensionless units.

The same regimes and almost the same dependencies were observed for other values of $n$, except $n = 1$. Note, that for real-life values of the dispersion coefficient $|d_2| \sim 0.005...0.5$ pump configuration with $n = 1$ does not provide large enough dispersion parameter for low pump threshold value.

## 4. DOPO in normal-dispersion photonic-crystal microresonator

### 4.1. Signal-mode Splitting

The threshold power reduction is an important task for practical applications. It will both reduce the impact of thermal and undesirable nonlinear effects and improve the energy efficiency of devices operating on the basis of the process of degenerate optical parametric oscillations. In [51] it was shown that it is possible to reduce the pump threshold value shifting particular microresonator modes, namely signal mode at $\mu = 0$ or side modes at $\mu = \pm 3n$.

There are several methods providing controllable mode shifts, including application of avoided mode crossings or photonic molecules [55-59]. In recent years, a method based on the use of nanostructured or photonic-crystal microresonators has been widely used. In photonic-crystal microring resonators the ring width can be modulated with the periodicity $\pi R / m$, where $R$ is the ring radius and $m$ is the azimuthal number of the targeted mode, to induce a bandgap at this mode. The induced bandgap results in the frequency splitting of the mode and one can control the magnitude of the mode splitting value by altering the modulation amplitude [16]. More complex modulation laws may provide simultaneous splitting of several modes [17,18,25,31].

In our work, we investigated the applicability of this approach for realization of the threshold optimization method proposed in [51]. We considered the case of the photonic-crystal microresonator when splitting coefficient is not the same for all modes.

First, we studied the case of the shifted central mode. For this purpose, we assumed $\beta_{\mu=0}=\beta^{s}$ for the signal mode ($\mu=0$) and $\beta_{\mu\neq0}=\beta^{bg}<<\beta^{s}$ for other modes [see Fig. 4].

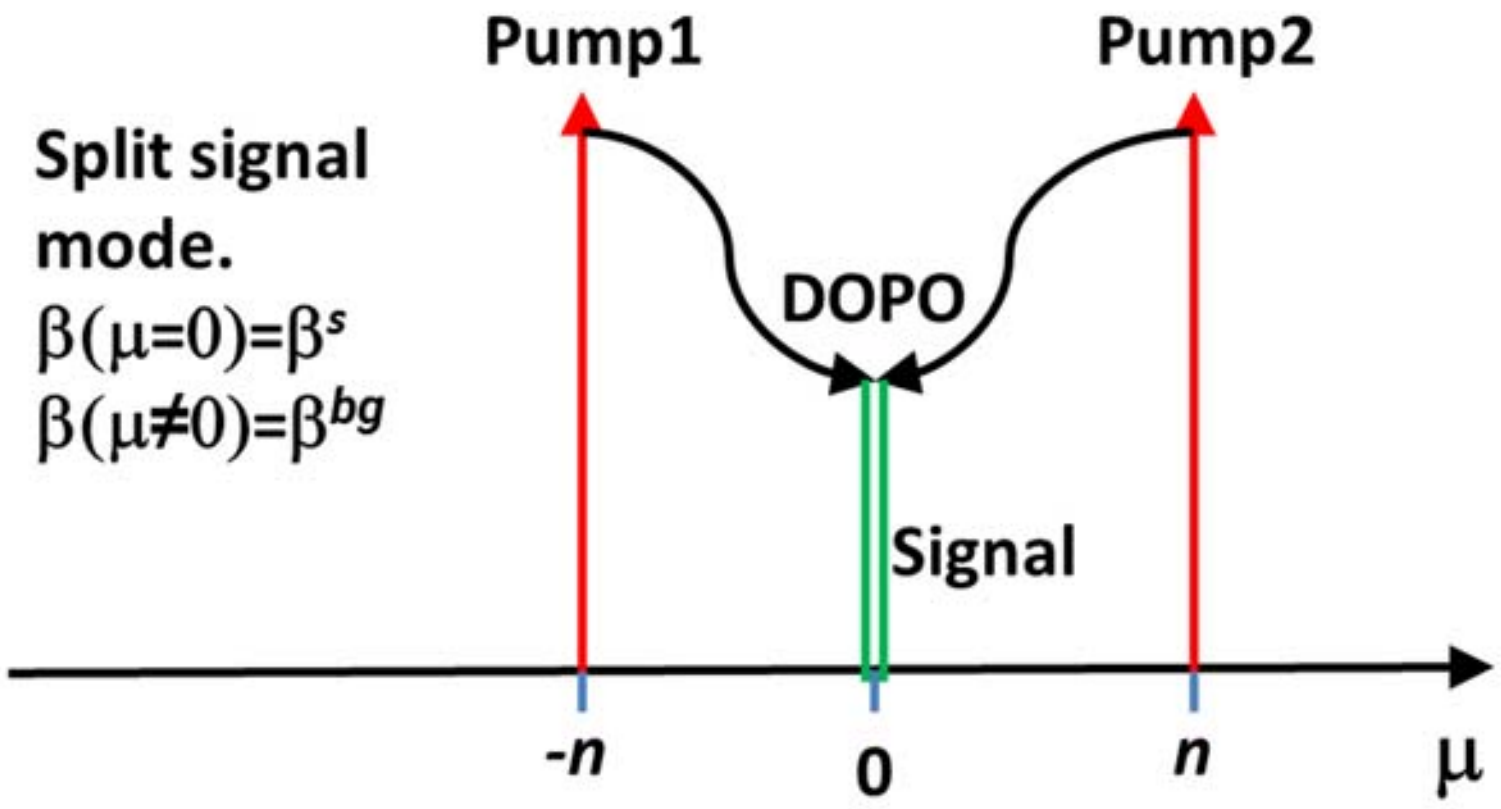

**Fig. 4.** Scheme of the process of DOPO at bichromatic pump in photonic-crystal microresonator with the split signal mode ($\beta_{\mu=0}=\beta^{s}, \beta_{\mu\neq0}=\beta^{bg}$). Double green line corresponds to the split mode.

It was revealed that introduction of the additional splitting of the signal mode by means of the photonic-crystal structure leads to the increase of the pump threshold [see Fig. 5]. Moreover, at sufficiently large splitting values, the process of generating the side frequencies at $\mu=\pm1$ becomes more probable than generation of the signal frequency at $\mu=0$. Besides the higher probability the side frequency generation requires smaller pump amplitude [see Figs. 5***(a)***, ***(b)***]. The reason of such effect may be the following: it was shown in [51] that threshold power reduction requires a particular direction of the signal mode shift. However, mode splitting simultaneously provides two variants of the shift, both positive and negative, and it is the non-optimal shift that plays the main role in the restructuring of the process dynamics.

Interestingly, if photonic-crystal structure provides signal mode splitting smaller than background splitting value, $\beta^s < \beta^{bg}$, pump threshold decreases in comparison with the structure with $\beta^s = \beta^{bg}$ and generation of non-desired components between pump lines, such as shown in Figs. 3***(b)***, ***(c)*** and which corresponds to areas 2, 3, 4 in Fig. 2, can be suppressed [see Figs. 5***(c)***, ***(d)***].

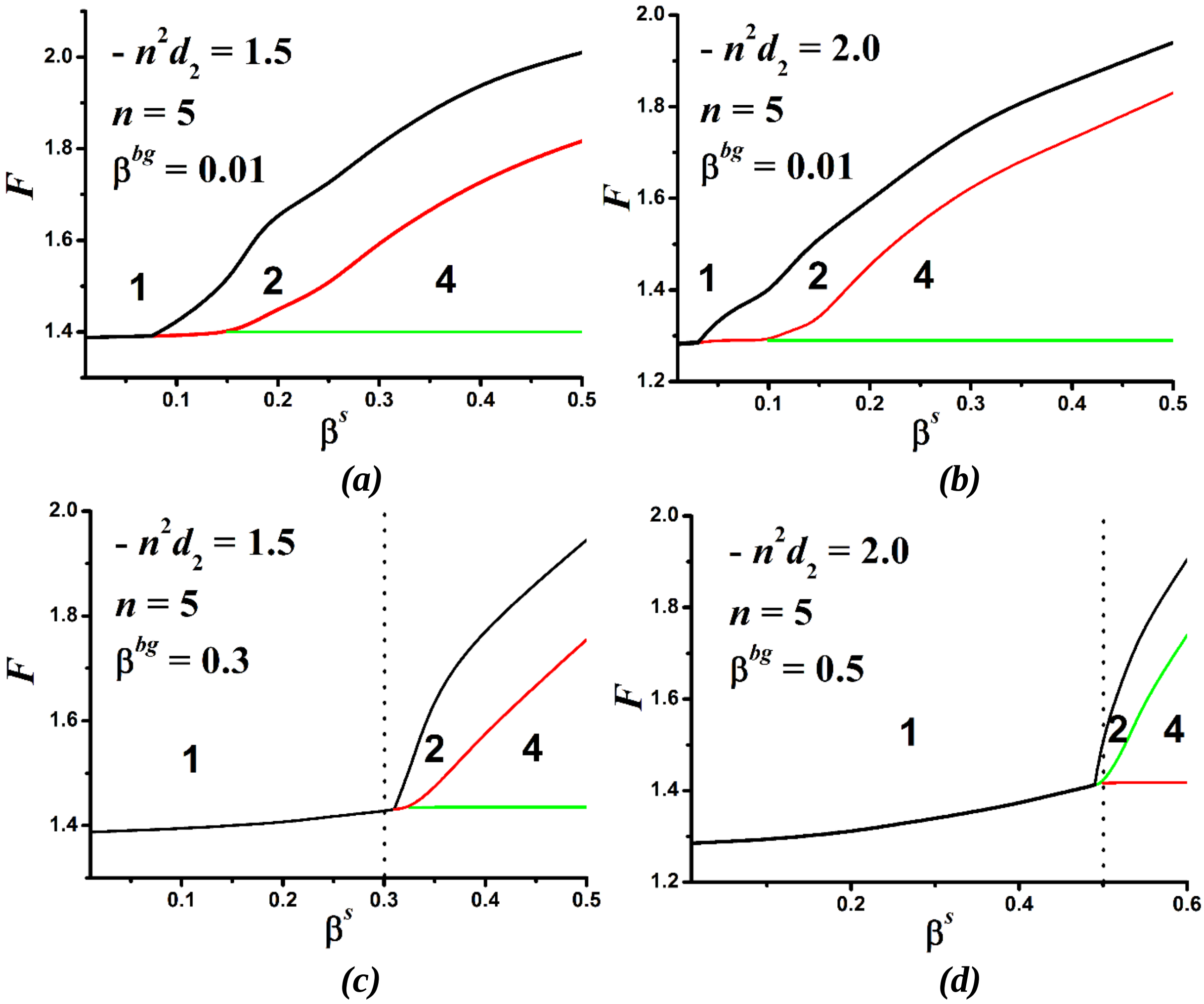

**Fig. 5.** Map of generation regimes arising in bichromatically-pumped normal dispersion photonic-crystal microresonator upon linear-in-time symmetric pump frequency scan for different values of the signal mode splitting $\beta^s$. Area's notations are the same as in Fig. 2. Dotted line in panels ***(c)*** and ***(d)*** corresponds to the case $\beta^s = \beta^{bg}$. In all cases $n = 5$, $\nu = 0.0005$. All quantities are plotted in dimensionless units.

### 4.2. 3*n*-Splitting

Also, it was shown in [51] that pump threshold can be diminished via frequency shift of side-modes at $\mu=\pm 3n$. Such approach can be realized in photonic-crystal microresonators with dual mode splitting [3n-splitting approach]. Such structures have been already demonstrated in the real-world photonic platforms, such as silicon nitride [25] or tantalum pentoxide ($Ta_2O_5$) [31].

For analysis of the applicability of this approach in photonic-crystal structures we used Eqs. (1) and set $\beta_{\mu=\pm 3n}=\beta^{3n}$ and $\beta_{\mu\neq\pm 3n}=\beta^{bg}\ll\beta^{3n}$ [see the scheme in Fig. 6].

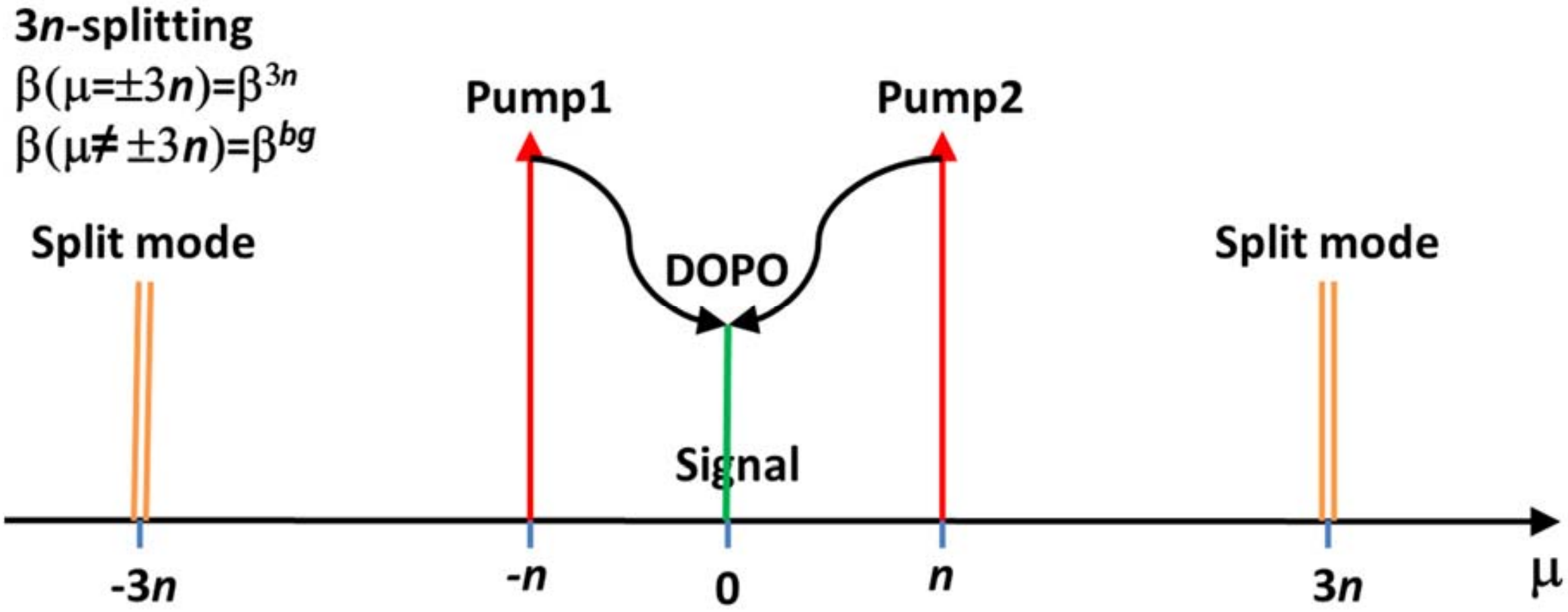

**Fig. 6.** Scheme of the process of DOPO at bichromatic pump at 3*n*-splitting in photonic-crystal microresonator with two split modes ($\beta_{\mu=\pm 3n}=\beta^{3n}, \beta_{\mu\neq\pm 3n}=\beta^{bg}$). Double lines correspond to the split modes.

It was revealed that the introduction of the splitting of side modes at $\mu=\pm 3n$ provides a decrease of the pump threshold. As shown in Fig. 7***(a)***, for each value of the dispersion parameter $-n^2 d_2$ pump threshold magnitude first decreases monotonically with the growth of the splitting value $\beta^{3n}$; however, when $\beta^{3n}$ exceeds some critical value depending on dispersion parameter pump threshold value begins to increase rapidly. We verified that the use of such structures preserves the required phase bistability property of the generated signal at $\mu=0$. Considered approach was found to be applicable for a wide range of dispersion parameter $-n^2 d_2$ and background splitting value $\beta^{bg}$ [see Fig. 7***(b)***]. As for the unmodulated structure, threshold value was found to increase with the growth of the background splitting value $\beta^{bg}$ [see Fig. 7***(b)***]. It was also revealed that larger value of $\beta^{bg}$ requires slightly larger optimal splitting values $\beta^{3n}$ induced by the photonic-crystal structure.

Also, similar to the case of unmodulated structure shown in Figs. 1 and 2 the optimized threshold value decreases as the dispersion parameter $-n^2 d_2$ increases [see Fig. 7***(a)***]. However, at large enough dispersion values [mostly for $-n^2 d_2 > 2.0$ and this critical value shifts to smaller values with the growth of $\beta^{bg}$] the same scenario of undesired nonlinear processes, e.g. non-deterministic generation of different signals between pump lines depending on the particular realization of the noise-like input conditions, shown in Section 3, also takes place [see maps of generation regimes for $-n^2 d_2 = 2.25$ and $-n^2 d_2 = 2.5$ in Figs. 7***(c)***, ***(d)***], that limits the effectiveness of using such structures at large values of the dispersion parameter.

Thus, for $\beta^{bg} = 0.01$ threshold value $F_{thr}$ can be reduced up to a value of the order of 1.0 for the dispersion parameter $-n^2 d_2$ close to 2.0 that requires splitting value $\beta^{3n}$ from 7.0 till 9.0. Note, that required splitting values are experimentally feasible in real-world microresonator systems, e.g. made of silicon nitride [25] or tantalum pentoxide ($Ta_2O_5$) [31].

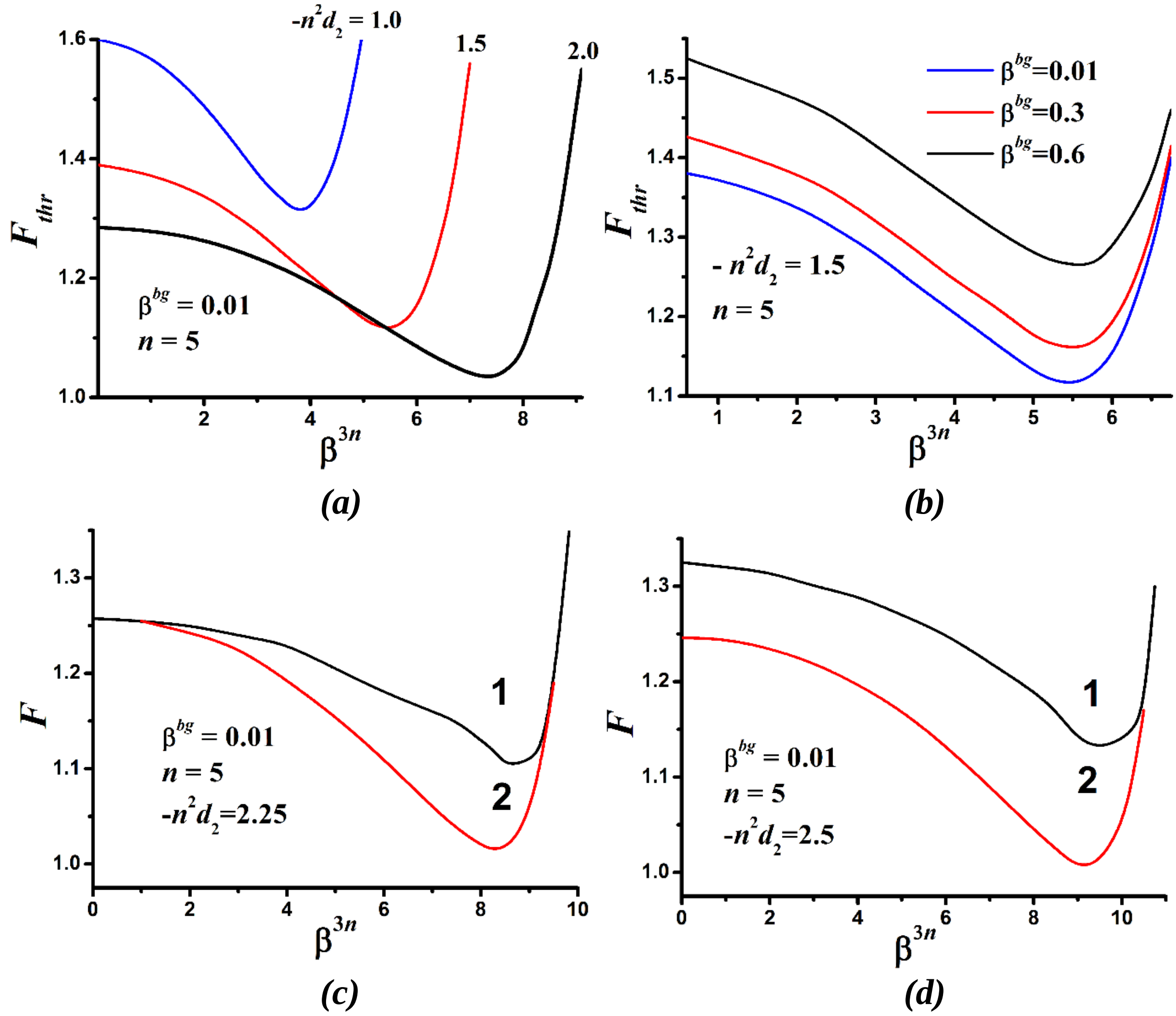

**Fig. 7.** ***(a,b)*** Pump amplitude threshold $F_{thr}$ vs splitting coefficient $\beta^{3n}$ for the process of the degenerate optical parametric oscillations in bichromatically-pumped normal-dispersion photonic-crystal microresonator with dual side modes splitting at $\mu = \pm 3n$ upon linear-in-time symmetric pump frequency scan for ***(a)*** different values of the dispersion parameter $-n^2 d_2$ and ***(b)*** different background splitting values $\beta^{bg}$. ***(c,d)*** Maps of generation regimes at ***(c)*** $-n^2 d_2 = 2.25$ and ***(d)*** $-n^2 d_2 = 2.5$ for different values of the splitting coefficient $\beta^{3n}$. Area's notations are the same as in Fig. 2. In all cases $n = 5$, $\nu = 0.0005$. All quantities are plotted in dimensionless units.

### 4.3. 5*n*-Splitting and 4-Modes Splitting

Note, that it is possible to use structures with larger interval between pumped and split modes, e.g. with splitting of side modes at $\mu = \pm 5n$ [see Fig. 8***(a)***; the splitting coefficient for such a structure is denoted as $\beta^{5n}$]. However, such structures require higher splitting values [compare Fig. 7***(a)*** and Fig. 8***(a)***]. For example, for mode splitting at $\mu = \pm 3n$ optimal splitting value for $-n^2 d_2 = 1.5$ and $\beta^{bg} = 0.01$ is about $\beta^{3n}_{opt} \approx 5.5$ and $F_{thr} \approx 1.115$ [see red line in Fig. 7***(a)***], while for the same parameters and mode splitting at $\mu = \pm 5n$ one has $\beta^{5n}_{opt} = 17.5$ and $F_{thr} \approx 1.195$ [see red line in Fig. 8***(a)***].

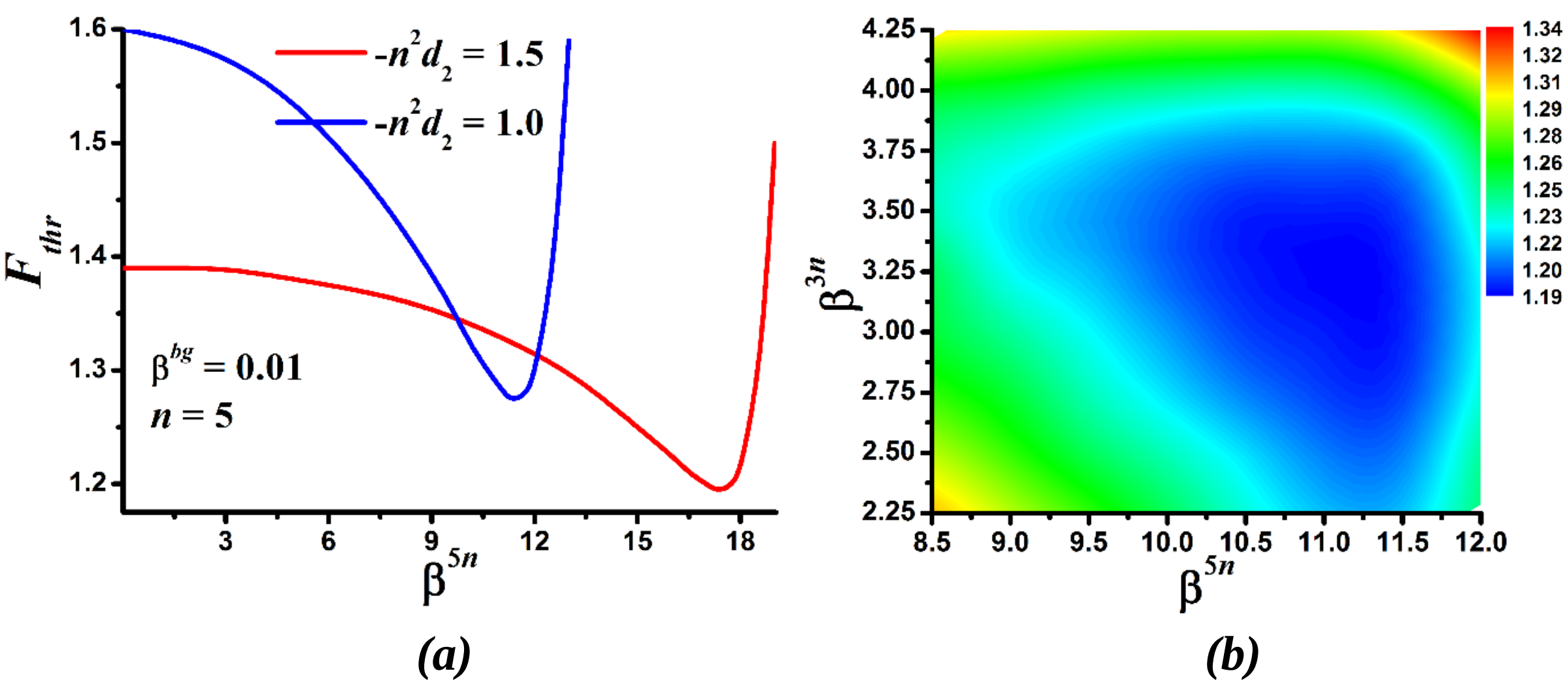

**Fig. 8.** ***(a)*** Pump amplitude threshold $F_{thr}$ vs splitting coefficient $\beta^{5n}$ for photonic-crystal microresonator with dual side modes splitting at $\mu = \pm 5n$ for different values of the dispersion parameter $-n^2 d_2$. ***(b)*** Pump amplitude threshold $F_{thr}$ vs splitting coefficient $\beta^{3n}$ at $\mu = \pm 3n$ and splitting coefficient $\beta^{5n}$ at $\mu = \pm 5n$ for photonic-crystal microresonator with simultaneous mode splitting at $\mu = \pm 3n$ and $\mu = \pm 5n$ at $-n^2 d_2 = 1.0$. In all cases $n = 5$, $\beta^{bg} = 0.01$, $v = 0.0005$. All quantities are plotted in dimensionless units.

It is possible to increase the efficiency of the pump threshold optimization using more complex photonic-crystal structures with simultaneous 4-modes splitting at $\mu = \pm 3n$ and $\mu = \pm 5n$ [see Fig. 9]. Optimal splitting values for such structures are close to those required for easier structure with mode splitting at $\mu = \pm 3n$ or at $\mu = \pm 5n$ discussed earlier. One may see in Fig. 8***(b)*** that if such method is applied at $-n^2 d_2 = 1.0$ and $\beta^{bg} = 0.01$ the threshold value is $F_{thr} \approx 1.19$, while splitting at $\mu = \pm 3n$ provides $F_{thr} \approx 1.315$ [see Fig. 7***(a)***] and the splitting at $\mu = \pm 5n$ results in $F_{thr} \approx 1.275$ [see blue line in Fig. 8***(a)***]. Thus, a system with 4-modes splitting provides an advantage over a system with only 2-modes splitting [3$n$- or 5$n$-splitting], but the difference is insignificant.

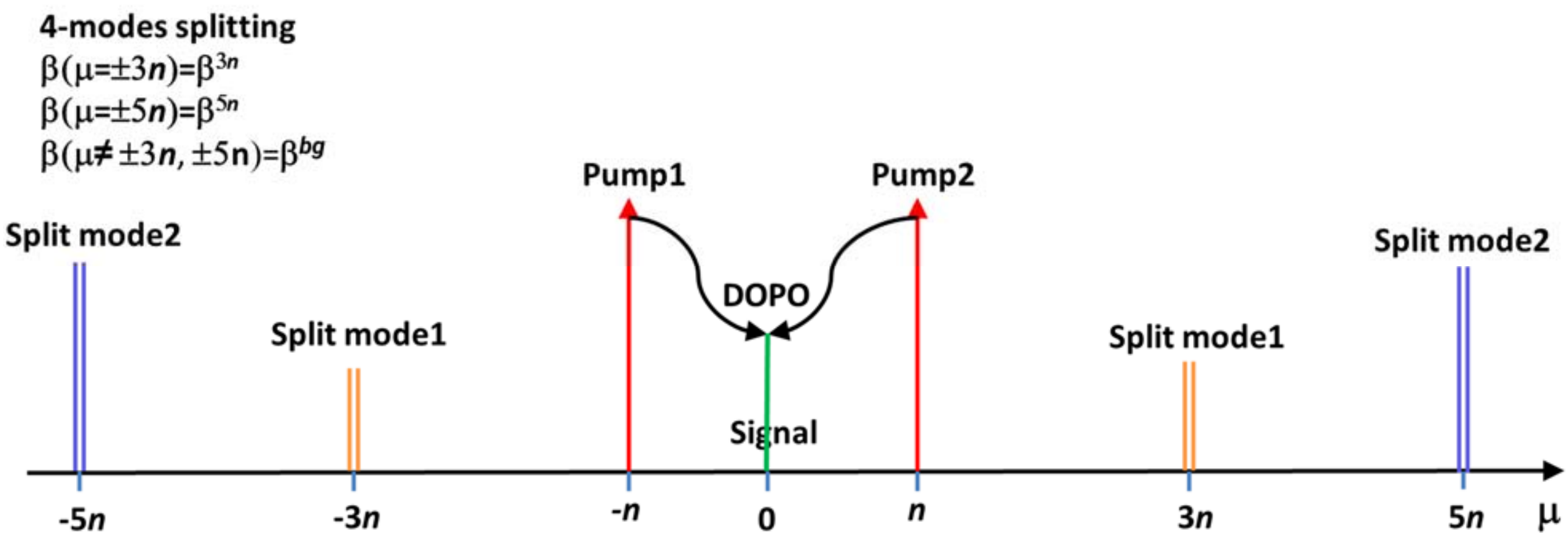

**Fig. 9.** Scheme of the process of DOPO at bichromatic pump in photonic-crystal microresonator with 4-modes splitting ($\beta_{\mu=\pm 3n} = \beta^{3n}, \beta_{\mu=\pm 5n} = \beta^{5n}, \beta_{\mu\neq\pm 3n,\pm 5n} = \beta^{bg}$). Double lines correspond to the split modes.

### 4.4. 2*n*-Splitting

Additionally to the methods proposed in [51], one more approach providing smaller threshold value for the desired DOPO process was found. Such result can be provided by photonic-crystal structure with mode splitting at $\mu = \pm 2n$ [2$n$-splitting approach, see Fig. 10].

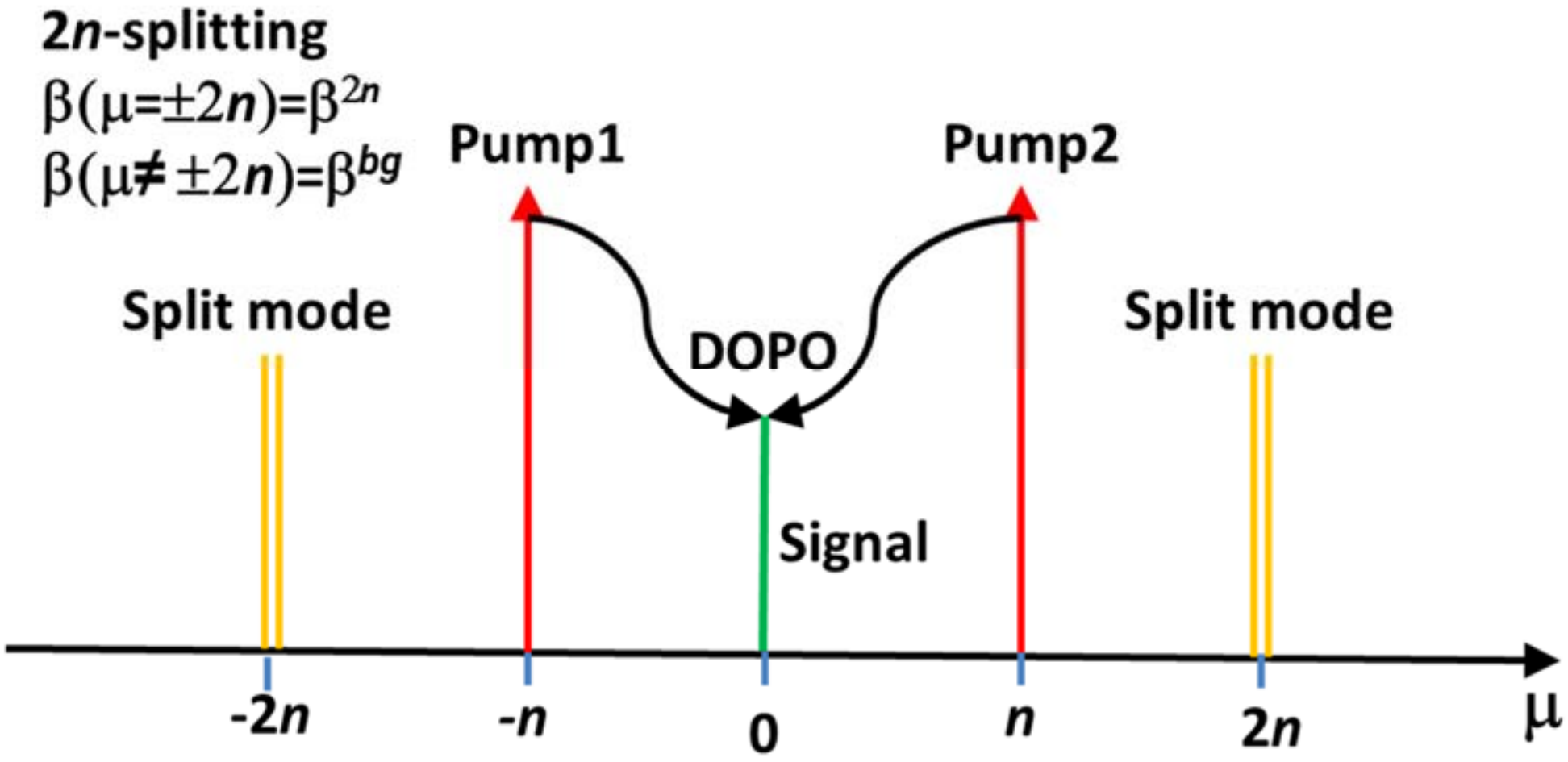

**Fig. 10.** Scheme of the process of DOPO at bichromatic pump at 2$n$-splitting in photonic-crystal microresonator with two split modes ($\beta_{\mu=\pm 2n} = \beta^{2n}, \beta_{\mu\neq\pm 2n} = \beta^{bg}$). Double lines correspond to the split modes.

It was revealed that above some critical splitting value $\beta_{cr}^{2n}$ dependent on the dispersion parameter degenerate optical parametric oscillations can be realized at significantly smaller pump amplitudes for a wide range of splitting value in comparison with $3n$-splitting approach [compare Fig. 11 and Fig. 7].

Note, that in this case the pump amplitude threshold value is less sensitive to the splitting value $\beta^{2n}$ [compare Fig. 7***(a)*** and Fig. 11***(a)***] and the dispersion parameter [compare Fig. 7***(a)*** and Figs. 11***(a), (b)***; $F_{thr_min}$ is minimal threshold value obtained for the particular dispersion parameter $-n^2 d_2$ and different splitting values $\beta^{2n}$] in comparison to structures with mode splitting at $\mu = \pm 3n$. Critical value of splitting $\beta_{cr}^{2n}$ was found to increase monotonically with the dispersion parameter $-n^2 d_2$ [see Fig. 11***(c)***] and to increase slightly with the background splitting parameter $\beta_0$. Significant reduction of the pump amplitude threshold can be observed for splitting values $\beta^{2n} > \beta_{cr}^{2n} + 0.5$ [see Figs. 11***(a)*** and 12]. Note, that for a wide range of the dispersion parameter required the splitting values are experimentally feasible in existing microresonator platforms, e.g. made in silicon nitride or tantalum pentoxide ($Ta_2O_5$). In considered structure if splitting becomes less than critical value the threshold jump occurs, and one may see the excitation of side modes at $\mu = \pm 1$ instead of signal generation at $\mu = 0$. For example, at $-n^2 d_2 = 2.5$ the critical value of splitting is $\beta_{cr}^{2n} \approx 4.23$. At $\beta^{2n} = 4.0$, that is less than critical value, one may observe only side modes excitation for pump amplitudes from $F = 1.25$ till at least $F = 1.5$. Above the critical value at $\beta^{2n} = 4.5$ signal mode excitation at $\mu = 0$ takes place for pump amplitudes from 0.73 till 1.15 [see Fig. 11***(d)***, blue line]. For pump amplitudes from 1.15 till 1.25 excitation of spectral components between pump lines is absent and for pump amplitudes from $F = 1.25$ till at least $F = 1.5$, only side modes excitation occurs.

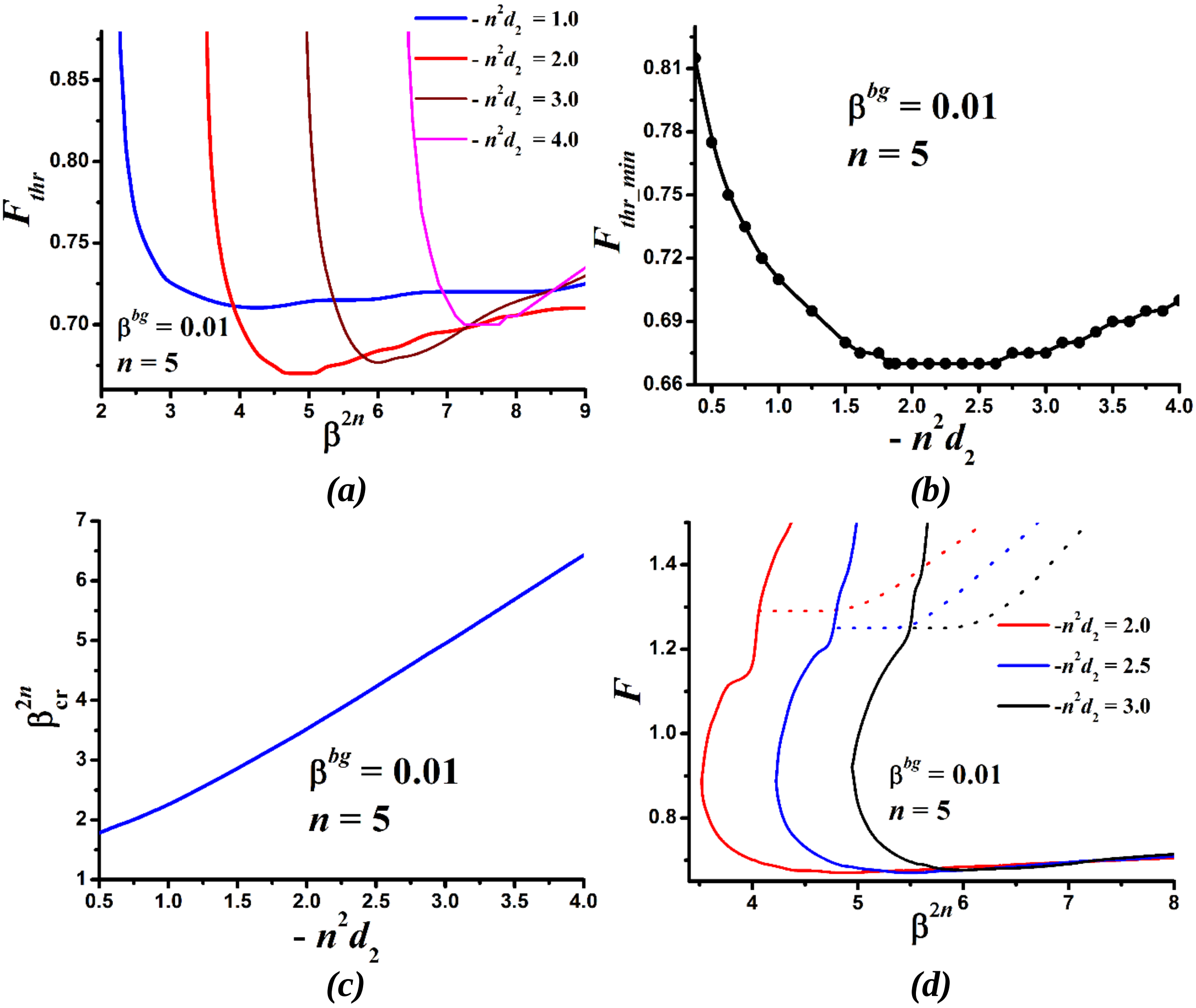

**Fig. 11.** ***(a)*** Pump amplitude threshold $F_{thr}$ vs splitting coefficient $\beta^{2n}$ for photonic-crystal microresonator with dual side modes splitting at $\mu = \pm 2n$ for different values of the dispersion parameter $-n^2 d_2$. ***(b)*** Minimal threshold value $F_{thr_min}$ obtained for the particular dispersion parameter $-n^2 d_2$ and different splitting values $\beta^{2n}$ vs dispersion parameter in the considered system. ***(c)*** Critical splitting value $\beta^{2n}_{cr}$ required for pump threshold optimization vs dispersion parameter. ***(d)*** Pump amplitude range applicable for DOPO realization. Solid lines of the same color describe upper and lower boundaries for the particular dispersion parameter (upper boundary exceeding 1.5 is not shown). Above the dotted and beyond the upper solid line signal generation at $\mu = 0$ is preceded by the generation of side modes at $\mu = \pm 1$. In all cases $n = 5$, $\beta^{bg} = 0.01$, $\nu = 0.0005$. All quantities are plotted in dimensionless units.

It's worth noting that in such a system the permissible pump amplitude $F$ may be limited from above. This is especially noticeable near the critical splitting value $\beta_{cr}^{2n}$ [see solid lines limiting applicable pump amplitude range in Fig. 11***(d)***; upper boundary exceeding 1.5 is not shown]. When this upper limit is exceeded, signal generation may either disappear completely or transform into side modes excitation. However, this limit rapidly increases with increasing splitting. At large splitting values and high enough pump amplitudes [amplitude range above dotted line and beyond upper solid line] during frequency scan one can first observe the excitation of sidebands at $\mu = \pm 1$, then transitioning to the excitation of the central line at $\mu = 0$. With the growth of the splitting value such regime shifts to larger pump amplitude values.

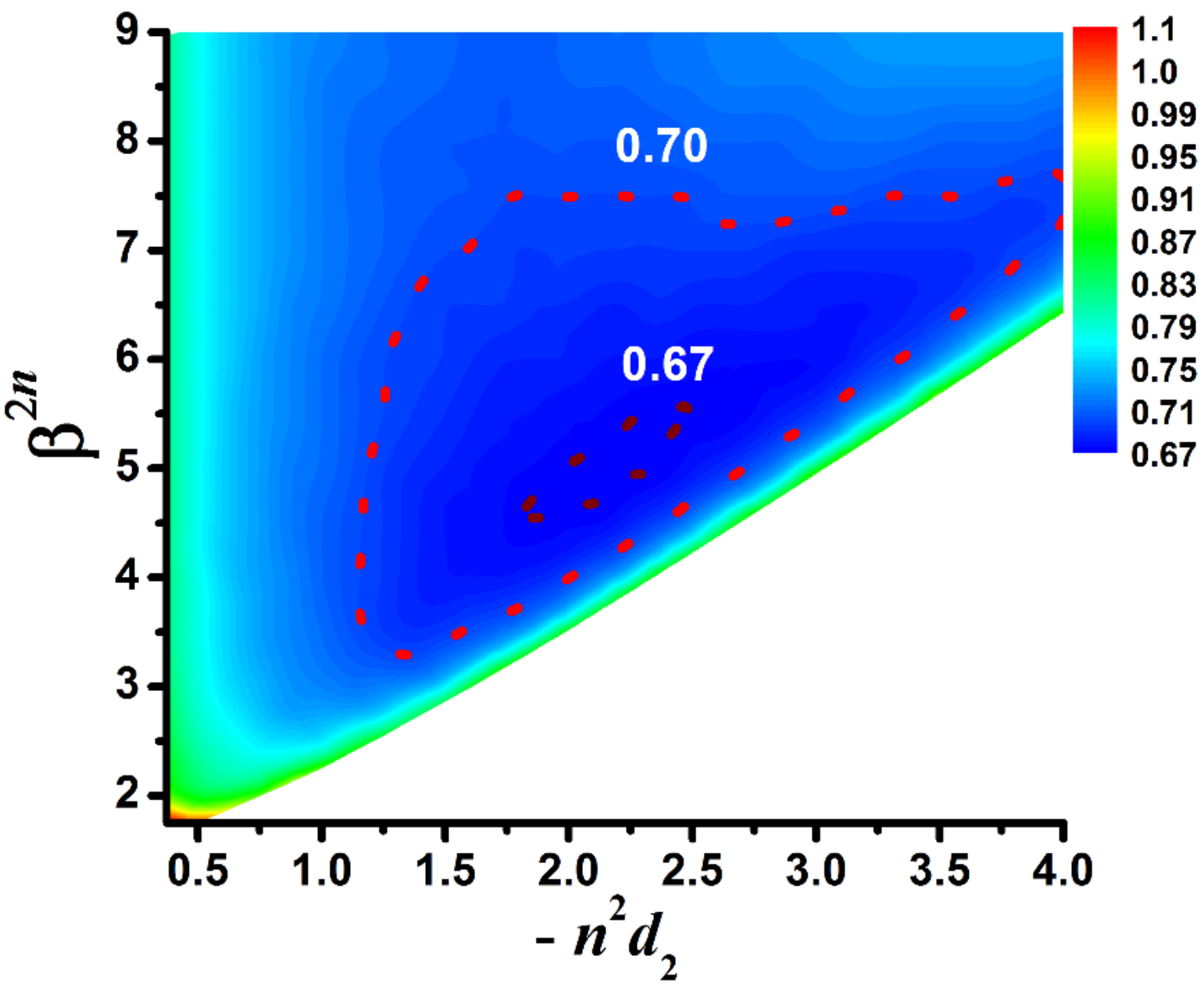

**Fig. 12.** Pump amplitude threshold $F_{thr}$ vs splitting coefficient $\beta^{2n}$ at $\mu = \pm 2n$ and dispersion parameter $-n^2 d_2$ for photonic-crystal microresonator with mode splitting at $\mu = \pm 2n$ ($2n$-splitting). Dotted lines bound the areas where threshold value is less than 0.67 and 0.7. Here $\beta^{bg} = 0.01$, $v = 0.0005$. All quantities are plotted in dimensionless units.

Achievable pump amplitude threshold for DOPO in such systems increases with the growth of the background splitting value $\beta^{bg}$. For example, at $-n^2 d_2 = 2.0$ $F_{thr_min}$ =0.67 for $\beta^{bg} = 0.01$, 0.73 for $\beta^{bg} = 0.3$, 0.84 for $\beta^{bg} = 0.5$.

It was revealed that simultaneous implementation of $2n$- and $3n$-splitting did not provide additional threshold reduction in comparison with $2n$-splitting.

We also checked that all described results can be obtained for different value of the interval between pump lines $2n$ and the considered process is insensitive to phase differences between the pump sources.

## 5. Discussion on Optimal Photonic-Crystal Structure for Low-Threshold DOPO

Thus, one may see that in unmodulated structure for $\beta^{bg} = 0.01$ the minimal pump amplitude threshold is about 1.24 at particular dispersion parameter [see Fig. 2]; $3n$-splittng approach provides pump threshold reduction up to 1.0, but in a particular range of the dispersion parameter and strict matching of this parameter and splitting value [see Fig. 7]. $2n$-splitting can lead to the threshold value up to 0.67 in small parameter range and to the value of 0.7 for the wide range of parameters [see Figs. 11 and 12].

We can assume that the mechanism for such a sharp decrease in the threshold is the following. The proposed photonic crystal structure enables the interaction of two processes, the considered process of the degenerate optical parametric oscillations via bichromatic pump and the mode-splitting-assisted process of optical parametric oscillations for each pump line [see Fig. 13]. Due to the presence of the mode shift at specific position, each pump line generates at least two sidebands, one of them corresponds to the signal frequency and acts as additional seeding for DOPO process. Such processes were described in normal-dispersion photonic-crystal microresonators in [30, 31]. Thus, such OPO process provides additional seeding for the process of the degenerate optical parametric oscillations resulting in a sharp decrease of its pump threshold [see Fig. 13].

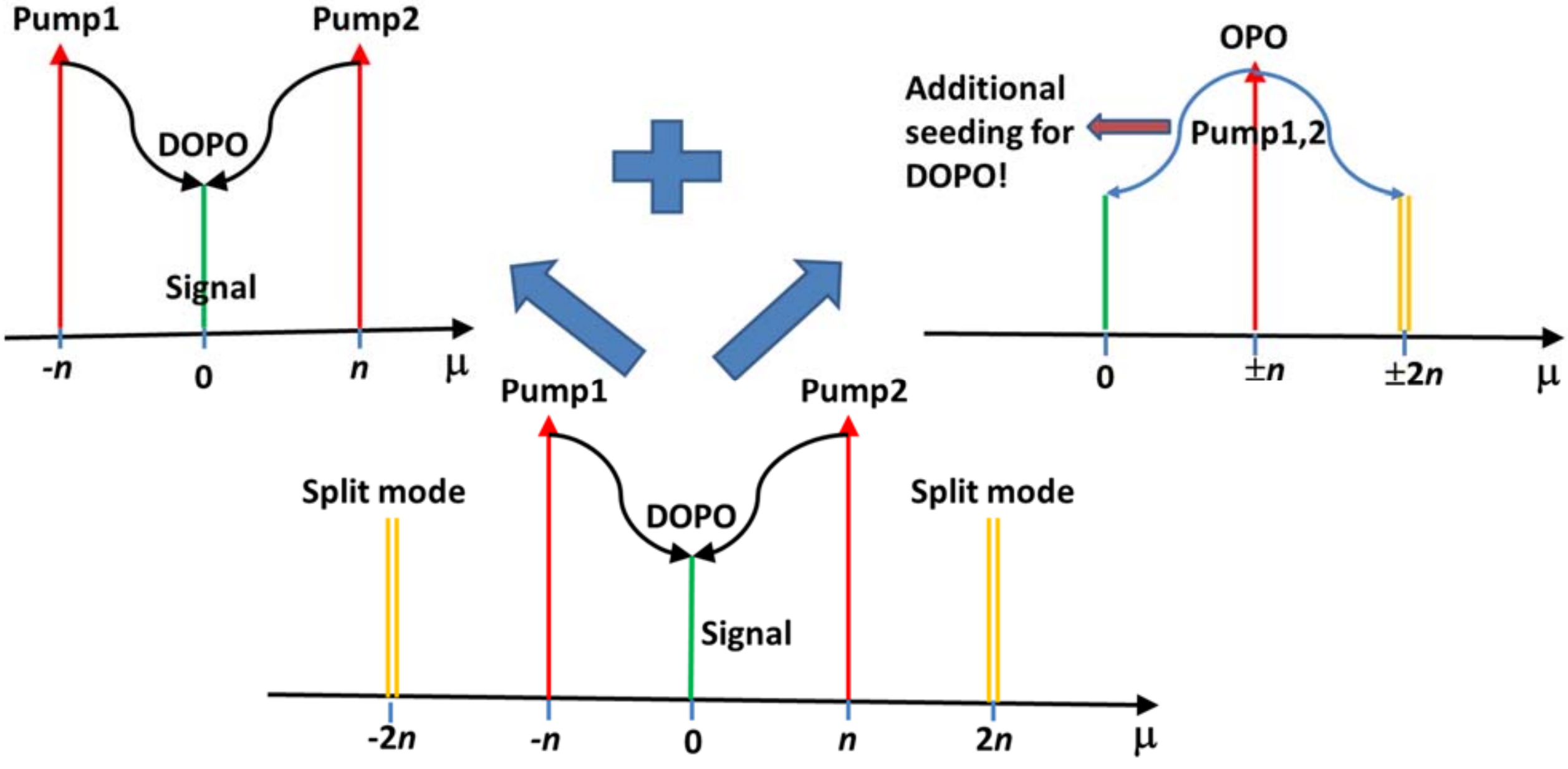

**Fig. 13.** Schematic description of the mechanism of DOPO threshold enhancement in photonic-crystal microresonator with $2n$-splitting.

## 6. Conclusion

The process of degenerate optical parametric oscillations via bichromatic pump is studied numerically in normal-dispersion photonic-crystal microresonator. It is demonstrated that photonic-crystal structure with two split modes located at particular interval from pumped modes provide significant decrease of pump power threshold for the considered process. The parameter range for this phenomenon is determined. It is revealed that optimal photonic-crystal structure leading to the minimal pump threshold value must provide splitting of modes at $\mu = \pm 2n$, while modes at $\mu = \pm n$ are pumped symmetrically. Introduction of the mode splitting at the signal mode leads to an increase in the generation threshold. The proposed method preserves the property of phase bistability of the generated signal, which is important for the creation of advanced computing systems. The obtained results provide better understanding of the dynamics of nonlinear processes in photonic-crystal microresonators and broaden the range of possible applications of such structures.

## Acknowledgments

This work is supported by the Russian Science Foundation (Project No. 25-12-00263).

The authors declare no competing interests.

## Data availability

The data that support the findings of this article are not publicly available. The data are available from the authors upon reasonable request.